\documentclass[preprint,12pt,aps,prd]{revtex4-1}
\usepackage{graphicx}
\usepackage{graphics}
\usepackage{subfigure}
\usepackage{verbatim}
 
\def\be{\begin{equation}}
\def\ee{\end{equation}}
\def\ba{\begin{array}}
\def\ea{\end{array}}
\def\bc{\begin{center}}
\def\ec{\end{center}}
\newcommand{\ra}{\rangle}
\newcommand{\la}{\langle}
\newcommand{\eq}{\begin{eqnarray}}
\newcommand{\en}{\end{eqnarray}}

\newcommand{\bfk}{{\bf k}_{\perp}}

\newcommand{\bfki}{{\bf k}_{\perp i}}
\newcommand{\bfb}{{\bf b}_{\perp}}
\newcommand{\bfbj}{{\bf b}_{\perp}^{j}}

\newcommand{\bfd}{{\bf \Delta}_{\perp}} 
\newcommand{\bfdj}{{\bf \Delta}_{\perp j }} 
\newcommand{\bfdi}{{\bf \Delta}_{\perp i}} 
\newcommand{\epsl}{\epsilon_\perp^{ij}}

\begin{document}
\title{Quark Wigner Distributions and GTMDs of Pion in the Light-Front Holographic Model}
\author{Navdeep Kaur and Harleen Dahiya}
\affiliation{Department of Physics, Dr. B. R. Ambedkar National Institute of Technology, Jalandhar-144011, India}
\begin{abstract}
We investigate the quark Wigner distributions of the pion to reveal the multidimensional picture of pion. We have used the spin improved wave functions of pion deduced from the light-front holographic model of mesons.
By using the Fock-state overlap representation, the Wigner distributions of an unpolarized, longitudinally polarized and transversely polarized quark inside the pion are calculated. We have presented the results of transverse Wigner distributions in impact-parameter space as well as in momentum space. In order to understand the role of skewness which gives the longitudinal momentum transfer between the quarks, we study the six-dimensional phase-space distribution: the generalized transverse momentum dependent parton distributions of pion for the case of zero skewness as well as for nonzero value of skewness.
\end{abstract}
\maketitle

\section{Introduction}
The understanding of hadronic structure in terms of quark and gluon degree of freedom is the main aim of hadronic physics for which a lot of theoretical and experimental efforts have been made in past years. The inclusive deep inelastic scattering provide such information through the parton distribution functions (PDFs). The PDFs however deliver information only about the longitudinal momentum fraction of partons (quarks and gluons) inside a hadron hence providing one dimensional picture of the hadron. A more comprehensive picture of hadrons is encoded in generalized parton distributions (GPDs) and transverse momentum dependent parton distribution functions (TMDs). The GPDs can be measured in hard exclusive reactions such as Deep virtual compton scattering (DVCS) and Deep virtual meson production (DVMP) giving the spatial distribution of partons in transverse plane along with the longitudinal momentum fraction of partons inside the hadron \cite{Ji:1997, Ji2:1997, Radyushkin:1997, Ji:1998, Ji2:1998, Blmlein:2000, Goeke:2001, Diehl:2003}. The TMDs however give details of transverse momentum distributions of partons inside the hadrons \cite{Tangerman95, Kotzinian95, Mulders97, Mulders98, Mulders07, Bacchetta:2008} through the semi-inclusive deep inelastic scattering (SIDIS) \cite{Bacchetta:2007, Ji:2005} and Drell-Yan (DY) processes \cite{Drell:1970, Christenson:1970, Collins:2002, Quintans:2011, Baranov:2014}. 

The complete picture of the hadronic structure can be obtained through the Wigner distributions which unify the position and momentum distributions \cite{Ji:2003, Belitsky:2004} and provide subtle details for partons inside the hadron. The five dimensional Wigner distributions are functions of impact-parameter, longitudinal momentum fraction and transverse momentum of partons. They have applications not only in high energy physics but in other areas such as signal analysis, quantum information, quantum molecular dynamics and heavy ion collisions \cite{Balazs:1984, Lee:1995, Hillery:1984}. The Winger distributions were introduced  in quantum chromodynamics (QCD) by Ji \cite{Ji:2003} and can be reduced to GPDs by integration over transverse momentum of parton and to TMDs by integration over impact-parameter. Through Fourier transformations, the Wigner distributions can be related to generalized transverse momentum distributions (GTMDs) which are the six-dimensional phase-space distributions also known as the mother distributions since they can be easily reduced to GPDs and TMDs. The GTMDs can directly describe the deep-inelastic lepton-nucleon scattering, virtual photon-nucleus quasi-elastic scattering and proton-nucleus collisions \cite{Hatti:2016, Zhon:2016, Ji:2017, Hatti:2017, Haqiwara:2017}. 

Many theoretical attempts have been made in various models to study the Wigner distributions of nucleons, for example, the light-front dressed quark model \cite{Mukherjee:2015, More:2017}, the light-cone spectator model \cite{Liu:2015}, the light-cone chiral quark soliton model \cite{Lorce:2012, Pasquini:2014,Lorce:2011}, the light-front quark-diquark model \cite{Satvir:2018} and the AdS/QCD inspired quark-diquark model \cite{Maji:2016, Chakrabarti:2016, Chakrabarti:2017}. Using Wigner distributions, the spin-orbital correlations can also be studied \cite{ Chakrabarti:2016, Satvir:2019}. However, the Wigner distributions of mesons have not been explored much in literature. Very recently, Wigner distributions of pion have been evaluated using light-cone quark model \cite{Lu:2018}.

The light-front framework, due to its simple light-front vacuum and light-front wave functions (LFWFs), is the ideal framework to describe the hadron structure.  The LFWFs carry information about the hadronic structure in terms of its constituents with their spin and orbital angular momentum and simply relates the constituent partons to their hadronic states. The knowledge of LFWFs allows one to study the different hadronic properties such as distribution amplitudes and structure functions. There exist several approaches through which the LFWFs can be extracted for hadrons, AdS/QCD correspondence being the most recent one. The holographic model constructed by AdS/QCD correspondence has several successful QCD applications like hadronic scattering processes, hadronic couplings \cite{Polchinski:2002, Janik:2000, Brodsky:2004, Levin:2009}, chiral symmetry breaking \cite{Rold:2005, Erlich:2005, Colangelo:2008}, hadronic spectrum \cite{Teramond:2005, Karch:2006, Forkel:2007}, quark potential \cite{Boschi:2006, Andreev:2006, Jugeau:} and hadron decays \cite{Hambye:2007}. The holographic Schr\"odinger equation for meson is an excellent equation to study the meson structure. Within the semiclassical approximation, one can recover the light-front wave functions for mesons from the solution of holographic Schr\"odinger equation \cite{Brodsky:2005, Brodsky:2006, Brodsky:2009, Brodsky:2014}. Since the dependence of the holographic LFWF on quark mass and helicity is necessary for phenomenological applications \cite{Swarnkar:2015}, one has to go beyond the semiclassical approximation where the quark masses can be included in LFWF \cite{Brodsky:2015}. In recent work \cite{Ahmady:2017, Ahmady:2018}, the quark spin has also been taken into account to provide better description of the experimental data on decay constants, charge radii, form factors etc. from Fermilab \cite{Patrignani:2016}.

In the light of the successes of the light-front holographic model, in the present work, we have studied the quark Wigner distribution of pion with minimal Fock state $ \vert q \bar{q} \rangle $. For this study we have used the spin-improved light-front wave function of light-front holographic model. At leading twist, the three Wigner distributions corresponding to the unpolarized, longitudinally polarized and transversely polarized quarks inside the pion are calculated by overlap representation of LFWFs. The results are discussed in the impact-parameter space as well as in the transverse space. Further, to understand the dependence of longitudinal momentum transfer from initial to final state of quark on the distributions, we have evaluated the twist-2 GTMDs of pion for zero skewness ($ \xi = 0 $) as well as for nonzero skewness ($ \xi \neq 0 $).  Four twist-2 GTMDs $ F_{1} $, $ G_{1} $, $ H_{1}^{k} $ and $ H_{1}^{\Delta} $ exist in this model for nonzero skewness. The $ H_{1}^{k} $ GTMD vanishes for zero skewness, therefore, we are left with three twist-2 GTMDs $ F_{1} $, $ G_{1} $ and $ H_{1}^{\Delta} $ in this model.

The manuscript is organized as follow: In Section II, we have detailed the light-front wave functions of pion in light-front holographic model. We have introduced the Wigner distribution of pion with different quark polarization and have presented the obtained numerical results of Wigner distribution in Section III. In Section IV, the GTMDs of pion for nonzero skewness as well as for zero skewness and their relation to Wigner distribution have been presented. We have summarized the obtained results in Section V.

\section{Light-front Holographic model}
Light-front holographic model is inspired by AdS/QCD correspondence which  connects the gravitational theory in five dimensional Anti-de Sitter (AdS) space-time to the Quantum Chromodynamics (QCD) in light-front formalism. A remarkable achievement of AdS/QCD correspondence is to provide the LFWFs for meson with valence Fock-state components which are known as the holographic light-front wave functions.
The holographic LFWFs of meson are obtained by holographic mapping between the dual field $ \Phi(\zeta) $ of AdS space and the LFWF $ \Psi (x,\zeta,\varphi) $ in physical space-time and are expressed as \cite{Brodsky:2014}
\eq
\Psi (x,\zeta,\varphi)= \frac{\phi(\zeta)}{\sqrt{2\pi \zeta}}\, X(x) \, e^{iL\varphi} \,,
\label{hwf}
\en
where $ X(x) = \sqrt{x(1-x)}$ is the longitudinal mode, $ \phi(\zeta) $ is the transverse mode and  $ \varphi $ is the angular dependence in the transverse light-front (LF) plane. Here $ \zeta^2 = x(1-x) \bfb^2$ is the transverse impact LF variable which provides the transverse separation of the quarks within the meson. The transverse mode $ \phi(\zeta) $ is related to the dual field $ \Phi(\zeta) $ of AdS space as
\eq
\phi(\zeta) = \zeta^{-3/2} \Phi(\zeta). 
\en
This transverse part of holographic LFWF for meson has been extracted from holographic LF Schr\"odinger equation as described in Refs. \cite{Brodsky:2011, Brodsky:2014}. With harmonic oscillator form of confining QCD potential, the solution of holographic light-front Schr\"odinger equation yields 
\eq
\phi_{nL}(\zeta) = \kappa^{1+L} \sqrt{\frac{2 n!}{(n+L)!}} \zeta^{1/2 + L} exp(\frac{-\kappa^2 \zeta^2}{2})L_n^L (\kappa^2 \zeta^2), 
\label{transverse_part_wf} 
\en
and meson mass spectrum
\eq
M^2 = 4 \kappa^4 \Big( n + L + \frac{S}{2} \Big).
\en
Here $ n $, $ L $ and $ S $ are the radial quantum number, internal orbital angular momentum and internal spin, respectively. The lowest possible solution exists for $ n = L = S = 0 $ which has been identified with the pion: a bound state of two massless quarks.
One can obtain the holographic light-front wave function for the pion by substituting Eq. (\ref{transverse_part_wf}) with $ L = S = 0 $ in Eq. (\ref{hwf}): 
\eq
\Psi (x,\zeta) &=& \frac{\kappa}{\sqrt{\pi}} \sqrt{x (1-x)}  \exp{ \left[ -{ \kappa^2 \zeta^2  \over 2} \right] } \,\nonumber\\
&=& \frac{\kappa}{\sqrt{\pi}} \sqrt{x (1-x)}  \exp{ \left[ -{ \kappa^2 x(1-x) \bfb^2  \over 2} \right] }, \,
\label{pionhwf} 
\en
for the case of massless quarks. Here $ \kappa $ is the AdS/QCD scale parameter.

By performing a two-dimensional Fourier transform of  Eq. (\ref{pionhwf}), the holographic LFWF in momentum space can be obtained as
\eq
 \psi_{\pi} (x,\bfk) = \frac{4 \pi N_{0}}{\kappa \sqrt{x (1-x)}}   \exp{ \left[ -\frac{\bfk^2 }{2 \kappa^2 x(1-x)} \right]}.
\label{pion-hwf-mom-space}
\en	
As the quark has nonzero mass, one can be include the quark masses by going beyond the semiclassical approximation as described in Ref.  \cite{Brodsky:2015}. 
The holographic LFWF with nonzero quark masses ($ m_u = m_d = m $) can be written as 
\eq
\psi_{\pi} (x,\bfk) = \frac{4 \pi N_{0}}{\kappa \sqrt{x (1-x)}}   \exp{ \left[ -\frac{\bfk^2}{2 \kappa^2 x(1-x)} \right]} \exp{ \left[ - \frac{m^2 }{2 \kappa^2 x (1-x) } \right]}.
\label{pion-hwf-quark-masses}
\en	
Here  $ N_0 $ is the normalization constant  satisfying  the following normalization condition  
 \eq
 	\int \mathrm{d}^2 \bfk \mathrm{d} x |\Psi(x,\bfk)|^2 = 1 \;. 
 	\label{norm}
 \en 
To obtain the helicity dependent holographic wave functions, the spin wave function has been included in the holographic LFWF of pion. The spin wave function for pseudoscalar pion results from the piontlike coupling of a pion with a $ q\bar{q} $ pair and is given by 
\eq
	S^{\pi}_{h \bar{h}} (x, \bfk)= \frac{\bar{u}_{h}(x,\bfk)}{\sqrt{\bar{x}}} \left[ \frac{M_\pi}{2P^+} \gamma^+ \gamma^5 +  \gamma^5 \right] \frac{v_{\bar{h}}(x,\bfk)}{\sqrt{x}}, 
\label{spin-structure} 
\en
where  $ M_{\pi} $ is the pion mass \cite{Ahmady:2018}. Here $ \bar{u}_{h}(x,\bfk) $ and $ v_{\bar{h}}(x,\bfk)$ represent the LF spinors of quark with helicity $ h $ and antiquark  with helicity $ \bar{h} $, respectively.

The complete picture of pion can be described by the total LFWFs which have been constructed from the product of a momentum-dependent wave function and a spin-dependent wave function. 
For two particle Fock state $ \vert q\bar{q} \rangle $ of a pion, the spin dependent LFWFs with different quark and antiquark helicities are expressed as 
 \eq 
\Psi^{\pi}_{+ \, +}(x,\bfk) &=& - \frac{k_{1} - i k_{2}}{x(1-x)}\, \psi_\pi (x,\bfk),\nonumber\\
\Psi^{\pi}_{+ \, -}(x,\bfk) &=& \Bigg(\frac{m}{x(1-x)}+M_{\pi} \Bigg) \, \psi_\pi (x,\bfk),\nonumber\\
\Psi^{\pi}_{- \, +}(x,\bfk) &=& \Bigg(-\frac{m}{x(1-x)}-M_{\pi} \Bigg) \, \psi_\pi (x,\bfk),\nonumber\\
\Psi^{\pi}_{- \, -} (x,\bfk) &=&- \frac{k_{1} + i k_{2}}{x(1-x)} \, \psi_\pi (x,\bfk),
 \label{pion-spin-hwf}
\en
where $ x $ and $ \bfk $ are the longitudinal momentum fraction and transverse momentum of active quark inside the pion, respectively.

\section{Wigner distribution of pion}
The Wigner distributions provide us the probabilistic information about hadrons. Even though the Wigner distributions do not provide us any direct information about the internal structure of hadrons but they can be reduced to  GPDs and TMDs after integrations over $ \bfb $ and $ \bfk $ respectively. 
Using the two-dimensional Fourier transforms of generalized quark-quark correlator $W^{[\Gamma]}(x, \bfk, \bfd)$ one can obtain the Wigner distributions $ \rho^{[\Gamma]}(x,\,\bfk, \bfb) $ of a quark as follows
\eq 
\rho^{[\Gamma]}(x,\,\bfk, \bfb) = \int\,\frac{\mathrm{d^2}\bfd}{(2\pi)^2}\, e^{[-i \, \bfb.\bfd]} \, W^{[\Gamma]}(x, \bfk, \bfd),
\label{wigner-distribution}
\en
where the generalized quark-quark correlator has the form
\eq
W^{[\Gamma]}(x,\,\bfk, \bf{\Delta_{\bot}}) &=& \frac{1}{2}\int\,\frac{\mathrm{d}z^{-} \mathrm{d^2}z_{\bot}}{(2\pi)^3}\, e^{[iP.z]}\,\la P^{''} \vert \bar\psi(-z/2)\Gamma \,{\cal W}\, \psi(z/2)\vert P^{\prime}\ra \Big\vert_{z^+ = 0}.
\label{wigner-correlator}
\en
Here $\Gamma$ ($ \gamma^+,  \gamma^+  \gamma^5, i \sigma^{j+}  \gamma^5 $) denote the Dirac $ \gamma $-matrices.  The averaged of the initial and final momentum of pion is denoted by $P = (P^{''} + P^\prime)/2$, $\Delta = P^{''} - P^\prime $ is the the momentum transfer to the pion, $k$ is the average momentum  and  $x= k^+/P^+$ is the longitudinal momentum fraction carried by the active quark. The presence of Wilson line $ \cal W $ ensures the color gauge invariance of the Wigner correlator.
  
As pion is a spin-0 meson, so in the present work to study the Wigner distributions we have considered the case of unpolarized pion only. Depending on various quark polarization configurations, the unpolarized pion has three Wigner distributions for a particular quark polarization.
The quark Wigner distributions for unpolarized (U), longitudinally polarized (L) and transversely polarized (T) quarks inside an unpolarized pion  can be defined as \cite{Lu:2018}  
\eq 
\rho_{UU}(x,\,\bfk, \bfb) &=& \rho^{[\gamma^+]}(x,\,\bfk, \bfb)\,, \nonumber\\
\rho_{UL}(x,\,\bfk, \bfb) &=& \rho^{[\gamma^+ \gamma^5]}(x,\,\bfk, \bfb)\,, \nonumber\\
\rho_{UT}(x,\,\bfk, \bfb) &=& \rho^{[i \sigma^j+ \gamma^5]}(x,\,\bfk, \bfb).
\en
The leading twist correlator function in terms of LFWFs can be expressed as
\eq
W^{[\gamma^+]}(x,\,\bfk, \bf{\Delta_{\bot}}) &=& {1\over 16\pi^3}\sum_{\lambda_{\bar{q}}} \left[\psi_{+\lambda_{\bar{q}}}^\star\left(x,\bfk^{o}
 \right)\psi_{+\lambda_{\bar{q}}}\left(x,\bfk^{i}
 \right)\right.\nonumber\\
&+&\left.\psi_{-\lambda_{\bar{q}}}^\star \left(x,\bfk^{o}
 \right)\psi_{-\lambda_{\bar{q}}}\left(x,\bfk^{i}
 \right) \right],
 \label{wigner-UU}  \\
W^{[\gamma^+ \gamma_5]}(x,\,\bfk, \bf{\Delta_{\bot}}) &=& {1\over 16\pi^3}\sum_{\lambda_{\bar{q}}}  \left[\psi_{+\lambda_{\bar{q}}}^\star\left(x,\bfk^{o}
 \right)\psi_{+\lambda_{\bar{q}}}\left(x,\bfk^{i}
 \right)\right.\nonumber\\
&-&\left.\psi_{-\lambda_{\bar{q}}}^\star\left(x,\bfk^{o}
 \right)\psi_{-\lambda_{\bar{q}}}\left(x,\bfk^{i}
 \right) \right],
\label{wigner-UL} \\
W^{[i\sigma^{j+}\gamma_5]}(x,\,\bfk, \bf{\Delta_{\bot}}) &=& {1\over 16\pi^3} \epsilon_{\bot}^{ji} \sum_{\lambda_{\bar{q}}}  \left[ (-i)^i \psi_{\uparrow\lambda_{\bar{q}}}^\star\left(x,\bfk^{o}
 \right)\psi_{\downarrow \lambda_{\bar{q}}}\left(x,\bfk^{i}
 \right)\right.\nonumber\\
&+& (i)^i \left.\psi_{\downarrow\lambda_{\bar{q}}}^\star\left(x,\bfk^{o}
 \right)\psi_{\uparrow \lambda_{\bar{q}}}\left(x,\bfk^{i}
 \right) \right],
\label{wigner-UT}
\en
 where the initial and final transverse momentum of active quark are
 \eq
 \quad \quad \bfk^{i} = \bfk + (1-x) \frac{\bfd}{2},  \quad \quad
 \bfk^{o} = \bfk - (1-x) \frac{\bfd}{2},  \nonumber
\en
respectively. Here $ \bfd $ represents the transverse momentum transfer. 
By using the LFWFs from Eq. (\ref{pion-spin-hwf}) in Eqs. (\ref{wigner-UU})-(\ref{wigner-UT}), we have obtained the following expressions for Wigner distribution of unpolarized quark inside an unpolarized pion, longitudinally polarized quark inside an unpolarized pion and transversely polarized quark inside an unpolarized pion represent as $\rho_{UU}$, $\rho_{UL}$ and $\rho_{UT}$ respectively
\eq
\rho_{UU}(x,\bfb,\bfk )&=&
  \frac{N_{0}^2}{\pi} \int \frac{d^2 \bfd} {(2\pi)^2}  e^{-i \bfd \cdot \bfb}\frac{2 \bfk^2 - (1-x)^2 \bfd^2 + 2 m^2 - x^2 (1-x)^2 M_\pi^2} {\kappa^2 x^3 (1-x)^3}\nonumber\\ 
&\times & \exp\left( -\frac{\bfk^2 + (1-x)^2 \frac{\bfd^2}{4}+ m^2}{ \kappa^2 x (1-x)}\right), 
\label{rho-UU} \\
\rho_{UL}(x,\bfb,\bfk ) &=& 
- \frac{N_{0}^2 (1-x)}{\pi} \int \frac{d^2 \bfd} {(2\pi)^2}  e^{-i \bfd \cdot \bfb}\frac{2 i \epsl \bfki \bfdj}{\kappa^2 x^3 (1-x)^3} \exp\left( -\frac{\bfk^2 + (1-x)^2 \frac{\bfd^2}{4}+ m^2}{\kappa^2 x (1-x)} \right),\nonumber\\
 \label{rho-UL}\\
 \rho_{UT}(x,\bfb,\bfk ) &=&
 -\frac{N_{0}^2 (1-x)}{\pi} \int \frac{d^2 \bfd} {(2\pi)^2}  e^{-i \bfd \cdot \bfb}\frac{i m \epsl  \bfdi}{\kappa^2 x^3 (1-x)^3} \exp\left( -\frac{\bfk^2 + (1-x)^2 \frac{\bfd^2}{4}+ m^2}{\kappa^2 x (1-x)} \right).\nonumber 
 \label{rho-UT}\\
\en
To obtain the purely transverse Wigner distributions, we have performed the integration over $ x $ as 
\eq
\rho_{UX}(\bfb,\bfk ) &=& \int\, dx \, \rho_{UX} (x,\bfb,\bfk ),
\en
where $ X $ stands for the polarization of quark inside unpolarized pion.

 \begin{figure*}
 \centering
 \includegraphics[width=0.4\columnwidth]{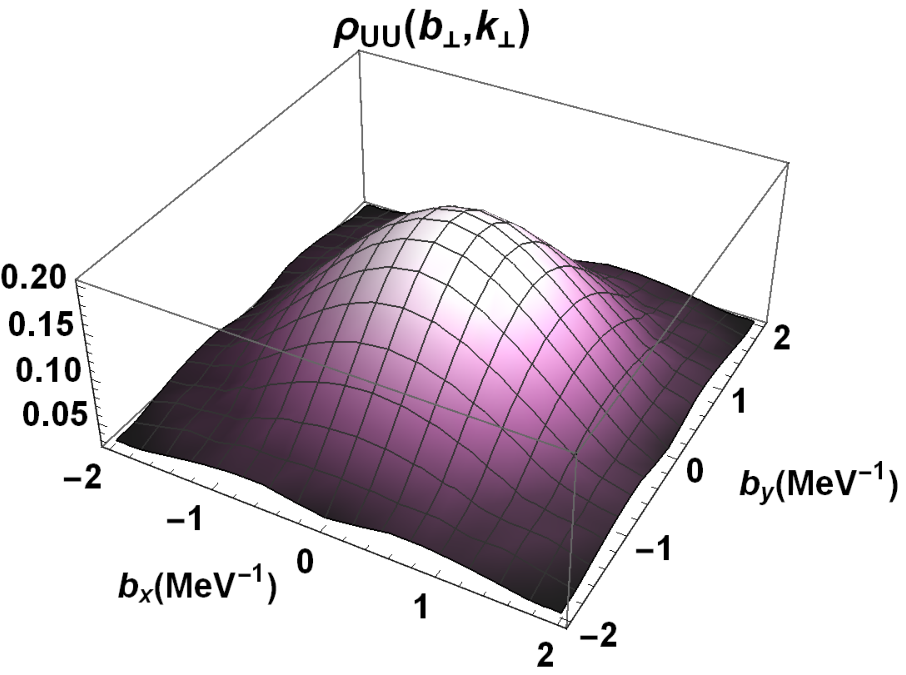}~~~~~~~
  \includegraphics[width=0.4\columnwidth]{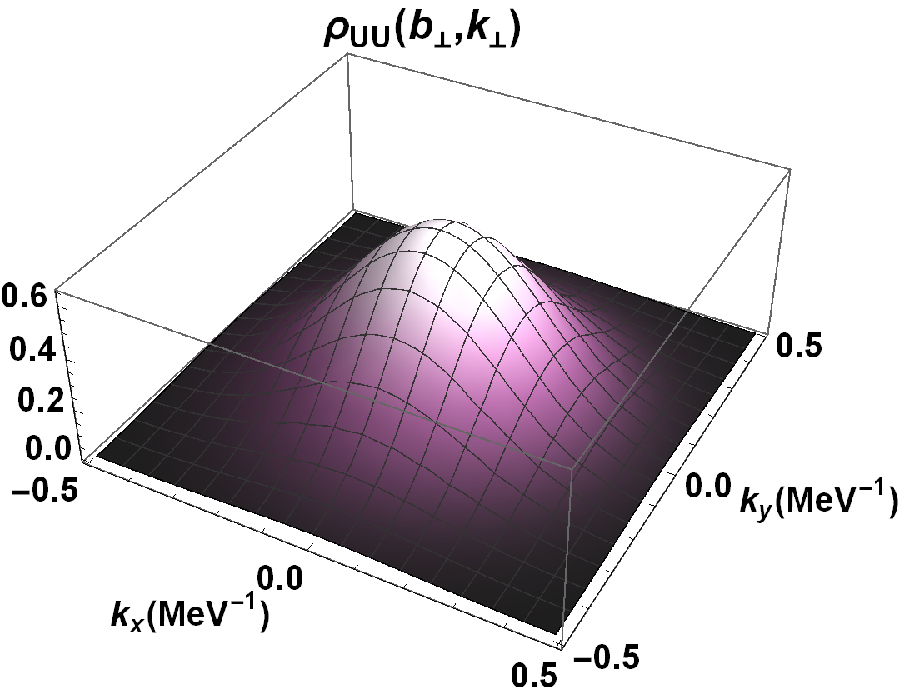}\\
\caption{Plot of unpolarized Wigner distribution for an unpolarized quark inside an unpolarized pion $\rho_{UU}(\bfb,\bfk )$ (a) in the impact-parameter space with fixed transverse momentum $\bfk = 0.3 $ GeV  (b) in the transverse-momentum space with fixed impact parameter  $\bfb = 0.3$ fm.}
  \label{plot_rho_UU}
\end{figure*}

The Wigner distributions depend on AdS/QCD scale parameter $\kappa$, mass of quark $m$ and mass of pion $ M_{\pi}$.  To obtain the numerical results we have used the following values of the parameters \cite{Ahmady:2018}
\[ \kappa = 523 ~ {\rm MeV},~~~~  m = 330~ {\rm MeV}~~~ {\rm and} ~~~  M_{\pi}=139 {\rm  MeV}. \]  
Wigner distribution of unpolarized quark inside the unpolarized pion, $\rho_{UU}(x,\bfb,\bfk )$ is  connected to the GPD $ H(x, \bfk) $ and to the unpolarized TMD of pion. In order to understand the dependence of  Wigner distribution on impact-parameter $ \bfb $ and transverse momentum $ \bfk $, in Fig. \ref{plot_rho_UU}, we have plotted the Wigner distribution of unpolarized quark inside the unpolarized pion $\rho_{UU}(\bfb,\bfk )$. The Wigner distribution  $\rho_{UU}(x,\bfb,\bfk )$ in the impact-parameter space with fixed transverse momentum  $\bfk = 0.3$ fm  is shown in Fig. \ref{plot_rho_UU} (a) whereas in Fig. \ref{plot_rho_UU} (b), the variation of Wigner distribution $\rho_{UU}(\bfb,\bfk )$ is shown in the transverse-momentum space with fixed impact parameter $\bfb = 0.3$ fm. $\rho_{UU}(\bfb,\bfk )$ has a circularly symmetric behavior in the impact-parameter space as well as in the transverse-momentum space which implies that the probability of the quark to flip up is equal to its probability to flip down. 
By comparing the distribution $\rho_{UU}$ in the impact-parameter space with $\rho_{UU}$ in the transverse-momentum space, we find that the peak of distribution is lower and also the distribution is more spread out in impact-parameter space as compared to that in transverse-momentum space.  This implies that distribution is more compact in transverse plane than in impact-parameter space. This will imply that the quark which is unpolarized and has larger transverse momentum inside the pion has smaller probability of being inside the pion.

\begin{figure*}
 \centering
  \includegraphics[width=0.4\columnwidth]{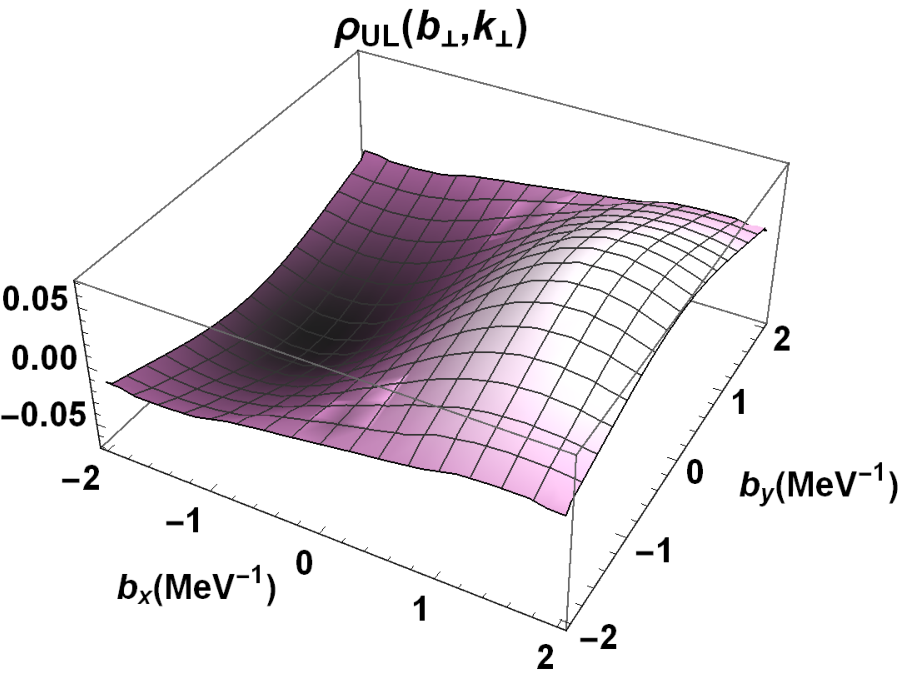}~~~~~~~
\includegraphics[width=0.4\columnwidth]{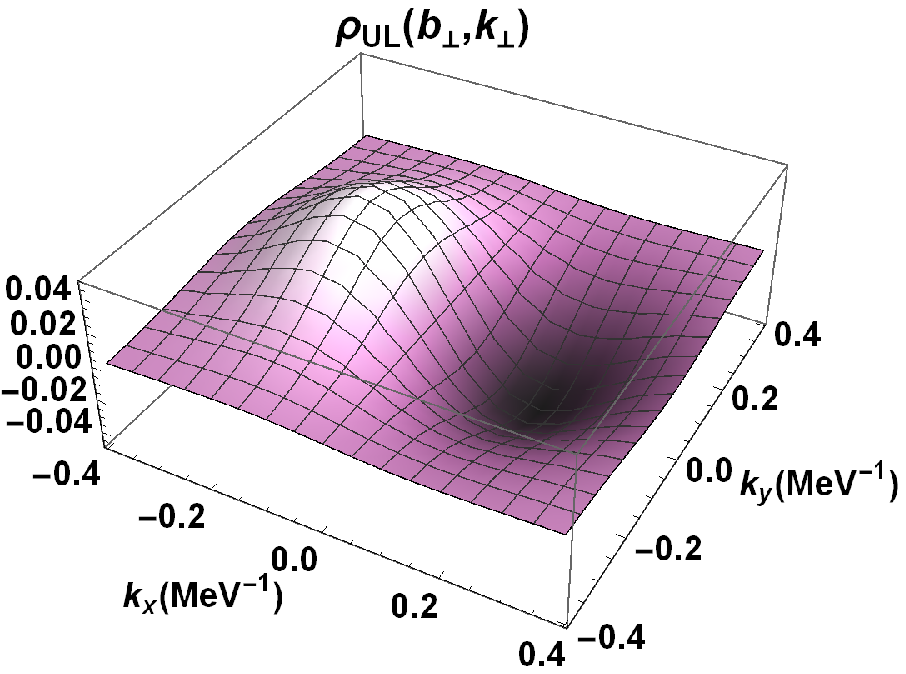}\\
\caption{Plot of longitudinal Wigner distribution for an longitudinally polarized quark inside an unpolarized pion $\rho_{UL}(\bfb,\bfk )$ (a) in the impact-parameter space with fixed transverse momentum $\bfk = 0.3 $ GeV  (b) in the transverse-momentum space with fixed impact parameter  $\bfb = 0.3$ fm.} \label{plot_rho_UL}
\end{figure*}

The Wigner distribution of a longitudinally polarized quark inside an unpolarized pion, $\rho_{UL}(\bfb,\bfk )$ has been shown in Fig.  \ref{plot_rho_UL}. In Fig. \ref{plot_rho_UL} (a), the longitudinal Wigner distribution  $\rho_{UL}(\bfb,\bfk )$ has been shown as a function of impact-parameter  with a fixed transverse momentum  $\bfk = 0.3$ fm whereas in Fig. \ref{plot_rho_UL} (b), we have shown $\rho_{UL}(\bfb,\bfk )$ as a function of transverse momentum  with fixed impact parameter $\bfb = 0.3$ fm. We observe that $\rho_{UL}(\bfb,\bfk )$ shows a dipolar structure in this model in both impact-parameter space as well as in the transverse-momentum space however,  the polarity is opposite in both spaces. In the impact-parameter space, the peak of distribution is around $ b_x = 1.5 $ fm whereas in the transverse-momentum space the peak is around $ k_x = 0.3 $ GeV. The privileged direction of quark polarization is responsible for the dipolar behavior of $\rho_{UL}(\bfb,\bfk )$. There is no GPD or TMD associated with this Wigner distribution $ \rho_{UL}(x, \bfb,\bfk) $ but it reflects the spin-orbital correlation. The correlation between quark spin and orbital angular momentum (OAM) in terms of Wigner distribution $ \rho_{UL}(x, \bfb,\bfk) $ can be written as \cite{Chakrabarti:2016, Chakrabarti:2017}
\eq
C_z^q=\int dx d^2\bfk d^2 \bfb (\bfb \times \bfk)_z \, \rho_{UL}(x, \bfb,\bfk).
\en
The quark spin is aligned parallel to the OAM if $ C_z^q > 0 $ and is aligned anti-parallel if $ C_z^q < 0 $.

\begin{figure*}
\centering
 \includegraphics[width=0.4\columnwidth]{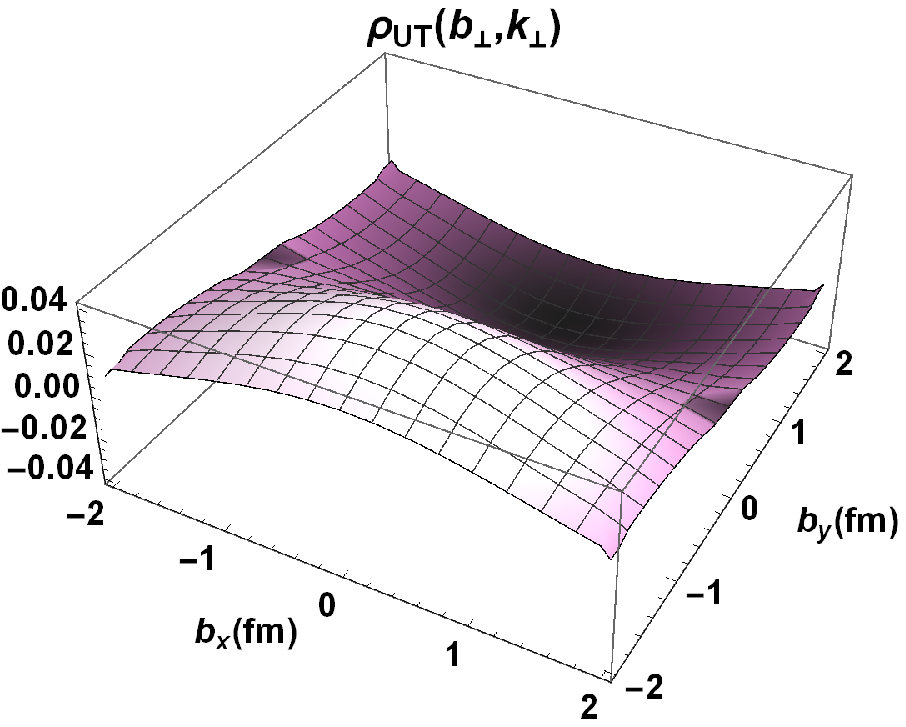}~~~~~~~
 \includegraphics[width=0.4\columnwidth]{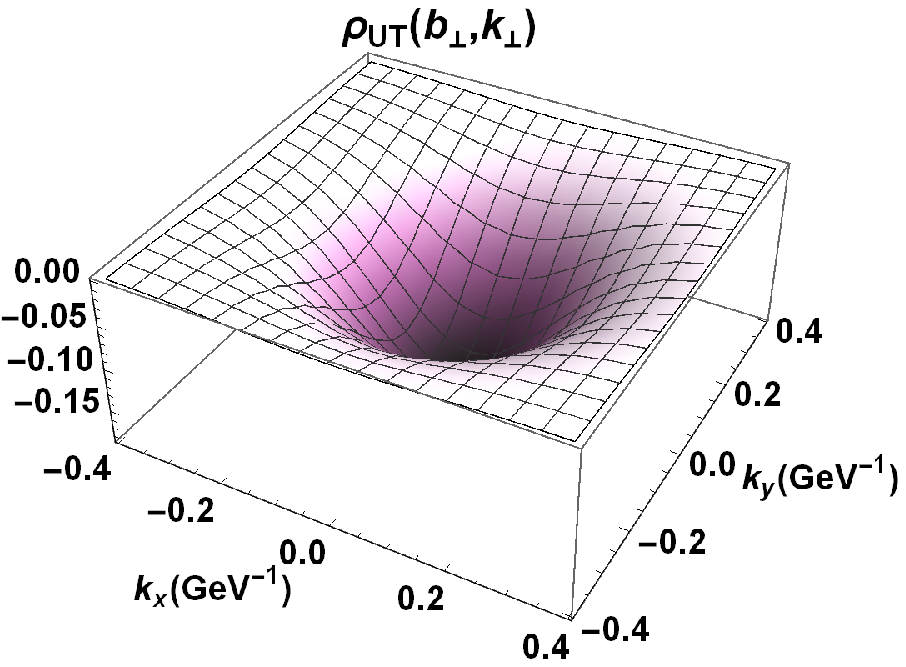}\\
\caption{Plot of transverse Wigner distribution for a transversely  polarized quark inside an unpolarized pion $\rho_{UT}(\bfb,\bfk )$ (a) in the impact-parameter space with fixed transverse momentum $\bfk = 0.3 $ GeV  (b) in the transverse-momentum space with fixed impact parameter  $\bfb = 0.3$ fm.}
 \label{plot_rho_UT}
\end{figure*}
In Fig. \ref{plot_rho_UT}, we have  presented the results for transverse Wigner distribution $\rho_{UT}(\bfb,\bfk )$ which describes the  distribution of polarized quark inside an unpolarized pion. For nonzero $\rho_{UT}(\bfb,\bfk )$ distribution, the polarization of quark must be perpendicular to the transverse co-ordinate because the distribution $\rho_{UT}(\bfb,\bfk )$ vanishes if it is in the same direction. This clearly indicates a relation between polarization direction and the transverse coordinate. In the  calculations, we have considered the polarization of quark along $x-$axis. The distribution $\rho_{UT}(\bfb,\bfk )$ shows a dipolar behavior in impact-parameter space as shown in Fig. \ref{plot_rho_UT} (a).  The polarity here depends on the polarization direction of quark. On the other hand, the distribution $\rho_{UT}(\bfb,\bfk )$ in momentum space has a symmetric behavior as shown in Fig. \ref{plot_rho_UT} (b). This symmetric behavior  conveys the non dependence of $\rho_{UT}(\bfb,\bfk )$ on the direction of transverse momentum of quark. The fact that the direction of polarization of quark is not related to transverse momentum of quark can be clearly seen from the expression of $\rho_{UT}(\bfb,\bfk )$ given in Eq. (\ref{rho-UL}). Therefore,  we can consider the direction of transverse momentum along any axis for quark polarization along $x-$axis. Further, by including the gluon contributions, one can relate the $\rho_{UT}(\bfb,\bfk )$ distribution to the T-odd GPD $ E_T $ and Boer-Mulder function $ h_1^{\bot} $.

\section{Generalized transverse momentum distributions (GTMDs)}
The study of GTMDs is crucial to  understand the role of skewness which is the fraction of longitudinal momentum carried by the quark. 
The leading twist GTMDs of pion are connected to the Wigner correlator given in Eq. (\ref{wigner-correlator}). We have 
\eq
W^{[\gamma^+]}(x,\,\bfk, \bf{\Delta_{\bot}}) &=&  F_{1} \vphantom{\frac{1}{1}}(x,\,\xi ,\bfk, \bfd)\,, \nonumber\\
W^{[\gamma^+ \gamma_5]}(x,\,\bfk , \bfd) &=&
 \frac{i\epsl k_\perp^i \Delta_\bot^j}{M^2} \, G_{1}^{k}(x,\,\xi ,\bfk, \bfd) \,,\nonumber\\
W^{[i\sigma^{j+}\gamma_5]}(x,\,\bfk, \bfd) &=&
 \frac{i\epsl k_\perp^i}{M} \, H_{1}^{k}(x,\,\xi ,\bfk, \bfd) + 
 \frac{i \epsl \Delta_\bot^i}{M} \, H_{1}^{\Delta}(x,\,\xi ,\bfk, \bfd) \,, \label{GTMDs}
\en
where $F_{1}$, $G_{1}^{k}$, $H_{1}^{k}$ and $H_{1}^{\Delta}$ are the leading twist GTMDs.  The transverse tensor $ \varepsilon_\perp^{ij} $ = $ \epsilon^{-+ij} $ has only two nonzero components i.e. $ \varepsilon_\perp^{12} = -\varepsilon_\perp^{21} = 1 $. The skewness parameter $ \xi $  provides information about the fraction of longitudinal momentum transfer between the initial and final state of quark. GTMDs can be studied for two different cases: one with $ \xi \neq 0 $ and other with $ \xi = 0 $.

The GTMDs of pion for zero skewness ($ \xi = 0 $) can be related to the Wigner distribution of pion as follow:
\eq
\rho_{UU}(x,\,\bfk, \bfb) &=& {\cal F}_{1}^{\pi} (x,\,\bfk, \bfb),\nonumber\\
\rho_{UL}(x,\,\bfk, \bfb) &=& \frac{\epsilon^{ij}}{M^2} \bfki \frac{\partial}{\partial \bfbj}{\cal G}_{1}^{\pi} (x,\,\bfk, \bfb),\nonumber\\
\rho_{UT}(x,\,\bfk, \bfb) &=& \frac{\epsilon^{ij}}{2 M} \bfki {\cal H}_{1}^{k} (x,\,\bfk, \bfb) + \frac{\epsilon^{ij}}{2 M} \frac{\partial}{\partial \bfbj}{\cal H}_{1}^{\Delta} (x,\,\bfk, \bfb).
\en
Since the longitudinal momentum carried by the quarks is nonzero in most of the experiments, it is necessary to study the distributions with nonzero skewness. As we are discussing the quark GTMDs here, we restrict ourselves to the DGLAP region with $ \xi < x < 1 $ corresponding to the quark distributions.

The overlap representation of the Wigner correlator to obtain GTMDs can be expressed as
\eq
W^{[\gamma^+]}(x,\,\bfk, \bf{\Delta_{\bot}}) &=& {1\over 16\pi^3}\sum_{\lambda_{\bar{q}}} \left[\psi_{+\lambda_{\bar{q}}}^\star\left(x^{out},\bfk^{out}
 \right)\psi_{+\lambda_{\bar{q}}}\left(x^{in},\bfk^{in}
 \right)\right.\nonumber\\
&+&\left.\psi_{-\lambda_{\bar{q}}}^\star \left(x^{out},\bfk^{out}
 \right)\psi_{-\lambda_{\bar{q}}}\left(x^{in},\bfk^{in}
 \right) \right],
 \nonumber \\
W^{[\gamma^+ \gamma_5]}(x,\,\bfk, \bf{\Delta_{\bot}}) &=& {1\over 16\pi^3}\sum_{\lambda_{\bar{q}}}  \left[\psi_{+\lambda_{\bar{q}}}^\star\left(x^{out},\bfk^{out}
 \right)\psi_{+\lambda_{\bar{q}}}\left(x^{in},\bfk^{in}
 \right)\right.\nonumber\\
&-&\left.\psi_{-\lambda_{\bar{q}}}^\star\left(x^{out},\bfk^{out}
 \right)\psi_{-\lambda_{\bar{q}}}\left(x^{in},\bfk^{in}
 \right) \right],
\nonumber\\
W^{[i\sigma^{j+}\gamma_5]}(x,\,\bfk, \bf{\Delta_{\bot}}) &=& {1\over 16\pi^3} \epsilon_{\bot}^{ji} \sum_{\lambda_{\bar{q}}}  \left[ (-i)^i \psi_{\uparrow\lambda_{\bar{q}}}^\star\left(x^{out},\bfk^{out}
 \right)\psi_{\uparrow\lambda_{\bar{q}}}\left(x^{in},\bfk^{in}
 \right)\right.\nonumber\\
& + & (i)^i \left.\psi_{\downarrow\lambda_{\bar{q}}}^\star\left(x^{out},\bfk^{out}
 \right)\psi_{\downarrow\lambda_{\bar{q}}}\left(x^{in},\bfk^{in}
 \right) \right],\label{wigner-GTMDs}
\en
 where 
 \eq
x^{in} &=& \frac{x+\xi}{1+\xi} , \quad \quad \bfk^{in} = \bfk + \frac{1-x}{1+\xi} \frac{\bfd}{2},  \nonumber
\en
and 
\eq
x^{out} &=& \frac{x-\xi}{1-\xi},\quad \quad \bfk^{out} = \bfk - \frac{1-x}{1-\xi} \frac{\bfd}{2}.  \nonumber
\en

Using Eqs. (\ref{pion-spin-hwf}), (\ref{GTMDs}) and (\ref{wigner-GTMDs}), the explicit expressions of the GTMDs obtained in this model are given below
\eq
 F_{1} (x,\,\xi ,\bfk, \bfd) &=& \frac{1}{16 \pi^3} \Bigg[2 \bfk^2 - \frac{(1-x)^2}{(1-\xi^2)} \frac{\bfd^2}{2} -\frac{2(1-x) \bfk \cdot \bfd}{(1-\xi^2)} \xi + 2 m^2  - \frac{M_\pi^2}{ {\cal Y}} \nonumber\\
&+& 2(1-x)m M_\pi \, {\cal X} \,\Bigg] \, {\cal Y} \,\, \psi_{\pi}^{\dagger} (x^{o},\bfk^{o}) \, \psi_\pi (x^{i},\bfk^{i}),\label{F1}\\
G_{1}^{k}(x,\,\xi ,\bfk, \bfd) &=& - \frac{2(1-x)}{16 \pi^3} \frac{M_\pi^2}{(1-\xi^2)} \, {\cal Y} \,\, \psi_{\pi}^{\dagger} (x^{o},\bfk^{o}) \, \psi_\pi (x^{i},\bfk^{i}),\label{G1}\\
H_{1}^{k}(x,\,\xi ,\bfk, \bfd) &=& \frac{2(1-x) M_\pi^2}{16 \pi^3} \, {\cal X}\, \, {\cal Y} \,\, \psi_{\pi}^{\dagger} (x^{o},\bfk^{o}) \, \psi_\pi (x^{i},\bfk^{i}),\label{HK}\\
H_{1}^{\Delta}(x,\,\xi ,\bfk, \bfd) &=& \frac{(1-x) M_\pi}{16 \pi^3} \Bigg[\frac{m}{(1-\xi)(1+\xi)} + \frac{M_\pi (1-x) (x-\xi^2)}{(1-\xi)^2 (1+\xi)^2}\Bigg]\, {\cal Y} \, \, \psi_{\pi}^{\dagger} (x^{o},\bfk^{o}) \, \psi_\pi (x^{i},\bfk^{i}).\nonumber\\
\label{HD}
\en
The terms $ {\cal X} $ and ${\cal Y}$ in the above equations are taken for the sake of simplification and are defined as
\eq
{\cal X} &=& \frac{(x - \xi)}{(1 - \xi)^2} -\frac{(x+\xi)}{(1+ \xi)^2}, \nonumber\\
{\cal Y} &=& \frac{(1 - \xi)^2 (1 + \xi)^2}{(x^2 - \xi^2)(1-x)^2} .
\label{XY}
\en

\begin{figure*}
  \centering
  \small{(a)}\includegraphics[width=0.4\columnwidth]{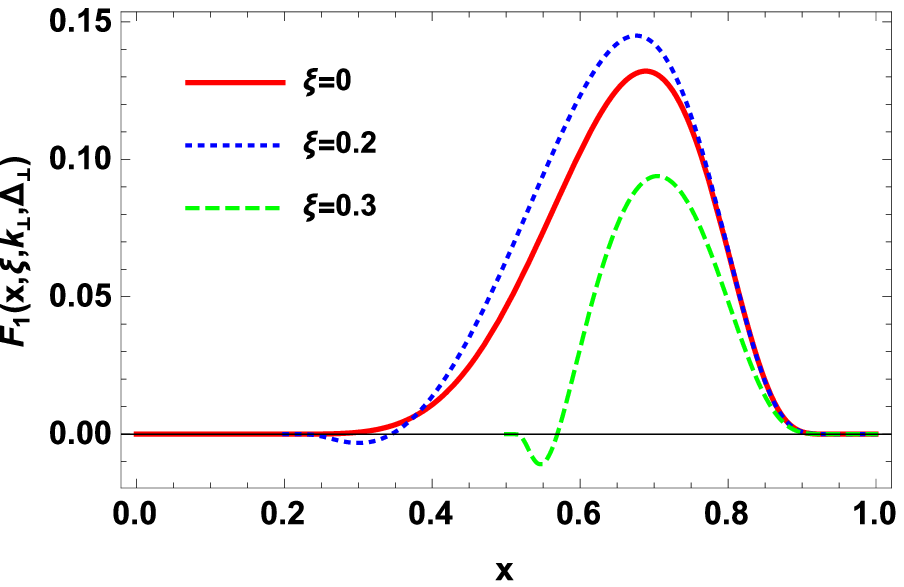}\hfill
  \small{(b)}\includegraphics[width=0.4\columnwidth]{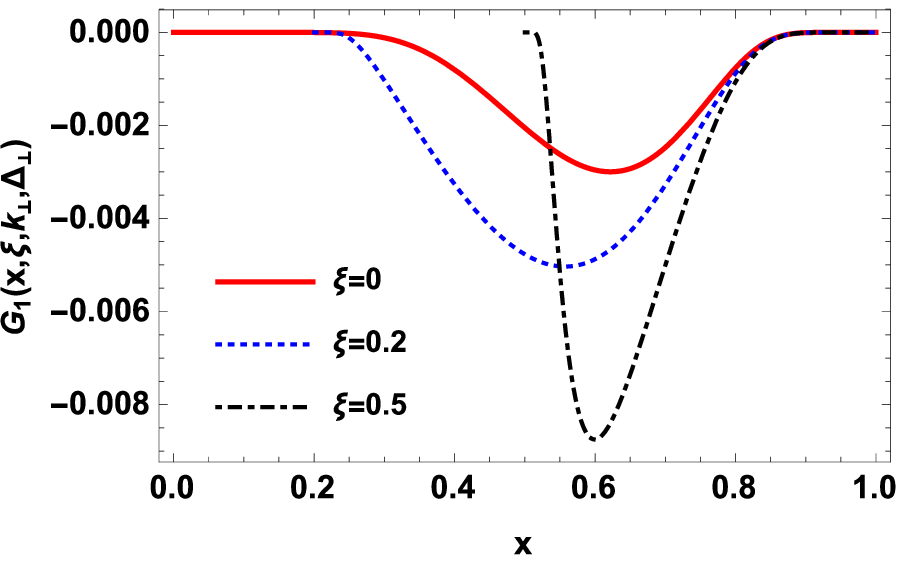}\\
  \small{(c)}\includegraphics[width=0.4\columnwidth]{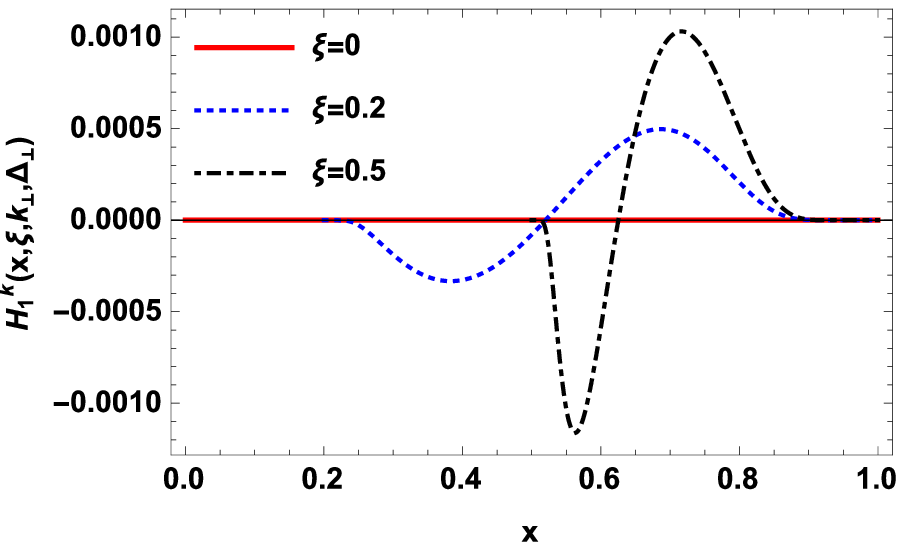}\hfill
  \small{(d)}\includegraphics[width=0.4\columnwidth]{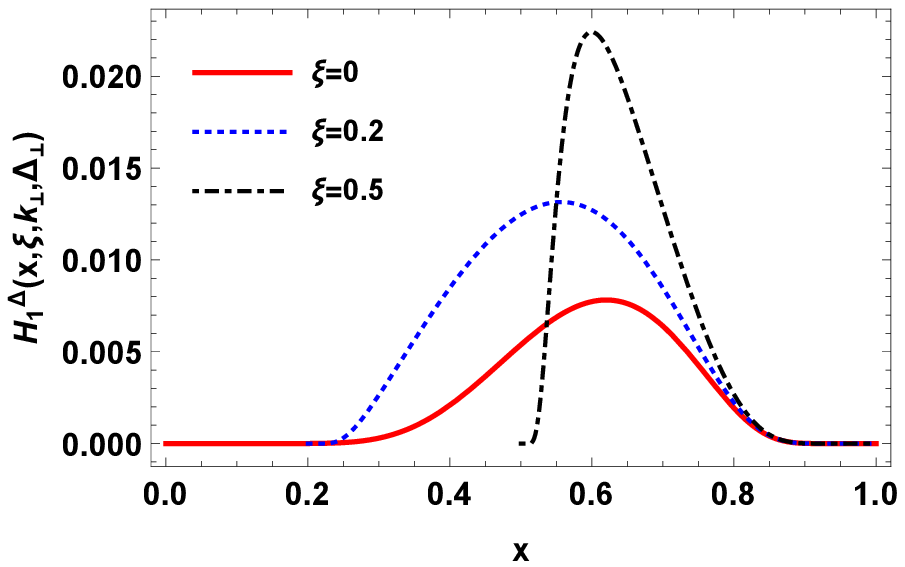}
  \caption{The quark GTMDs $ F_{1} (x,\,\xi ,\bfk, \bfd) $, $ G_{1}(x,\,\xi ,\bfk, \bfd) $, $ H_{1}^{k} (x,\,\xi ,\bfk, \bfd) $ and $ H_{1}^{\Delta} (x,\,\xi ,\bfk, \bfd) $ plotted as a function of $ x $ for different values of $ \xi $ including  $ \xi = 0 $. The values of $ \bfk$ and $ \bfd$ have been fixed as $ \bfk = 0.5 $ GeV and $ \bfd = 1.5 $ GeV .}
 \label{plot_F1_G1}
\end{figure*}

\begin{figure*}
\centering
\begin{minipage}[c]{1\textwidth}
(a)\includegraphics[width=.4\textwidth]{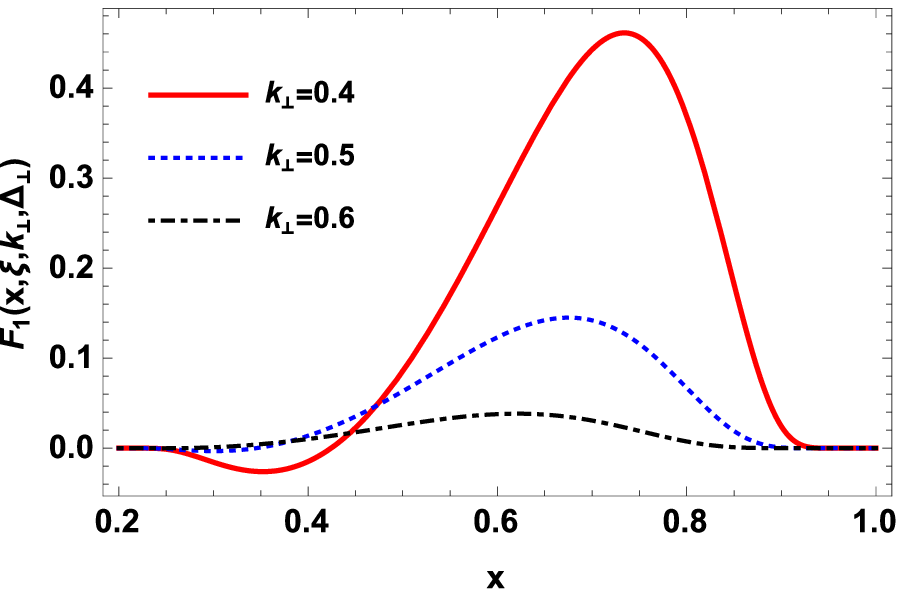}\hfill
(b)\includegraphics[width=.4\textwidth]{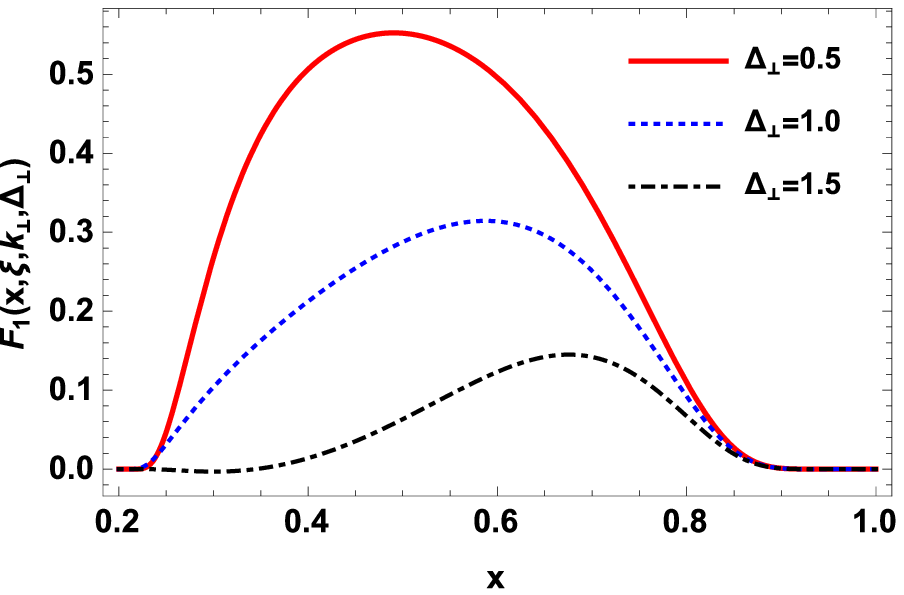}
\end{minipage}
\begin{minipage}[c]{1\textwidth}
(c)\includegraphics[width=.4\textwidth]{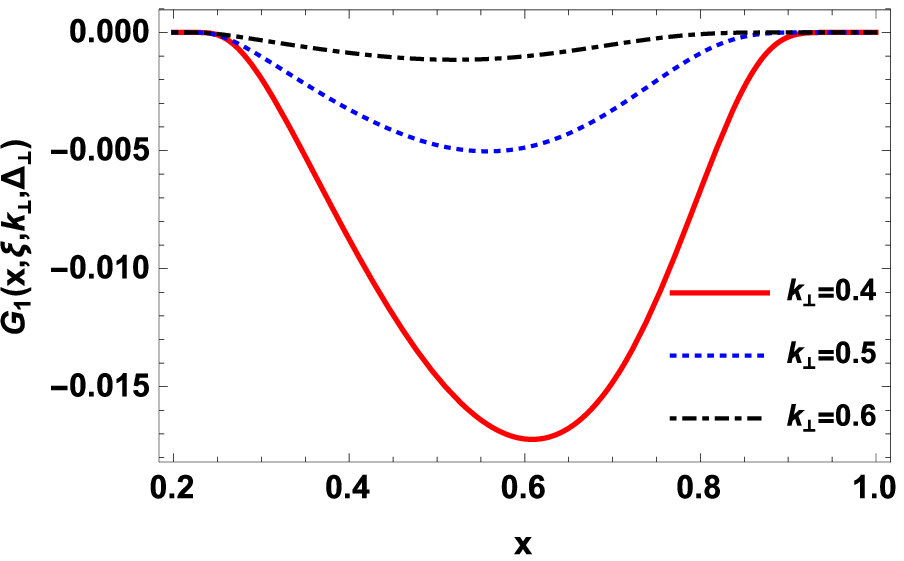}\hfill
(d)\includegraphics[width=.4\textwidth]{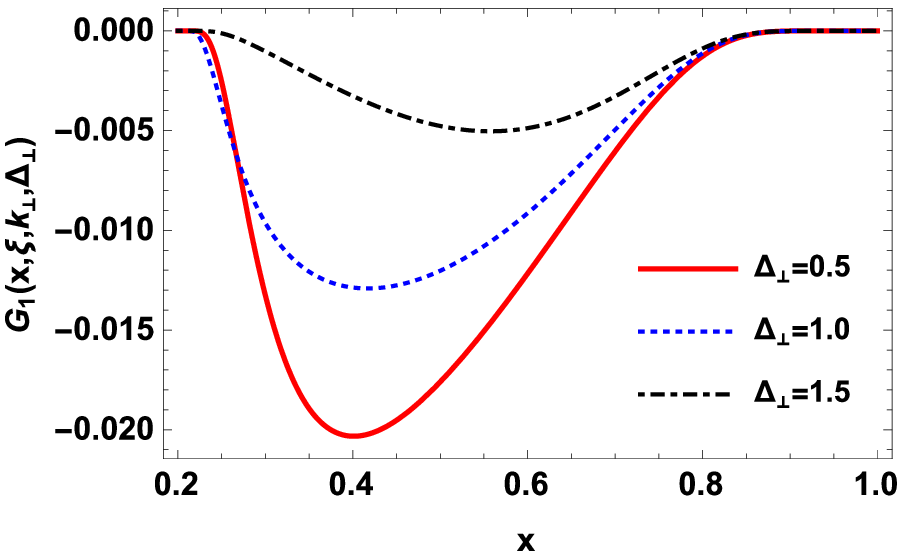}
\end{minipage}
\begin{minipage}[c]{1\textwidth}
(e)\includegraphics[width=.4\textwidth]{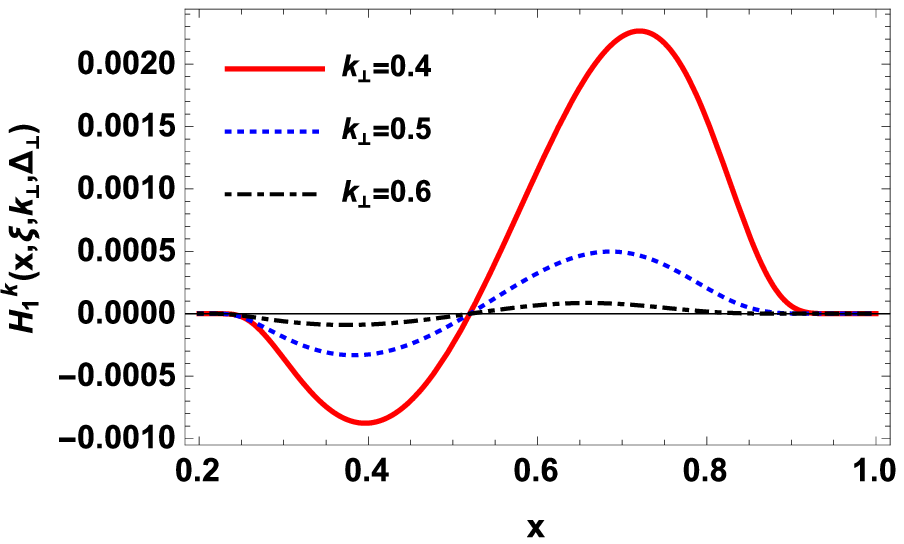}\hfill
(f)\includegraphics[width=.4\textwidth]{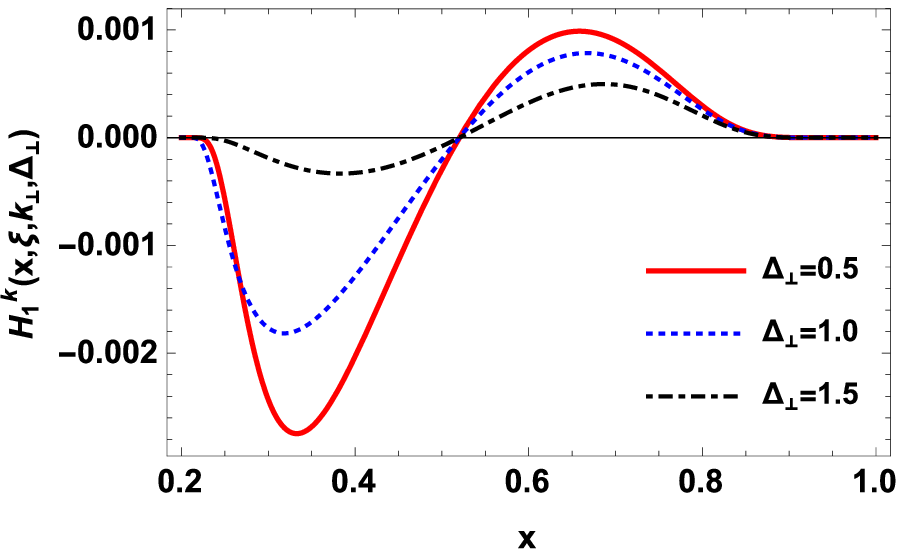}
\end{minipage}
\begin{minipage}[c]{1\textwidth}
(g)\includegraphics[width=.4\textwidth]{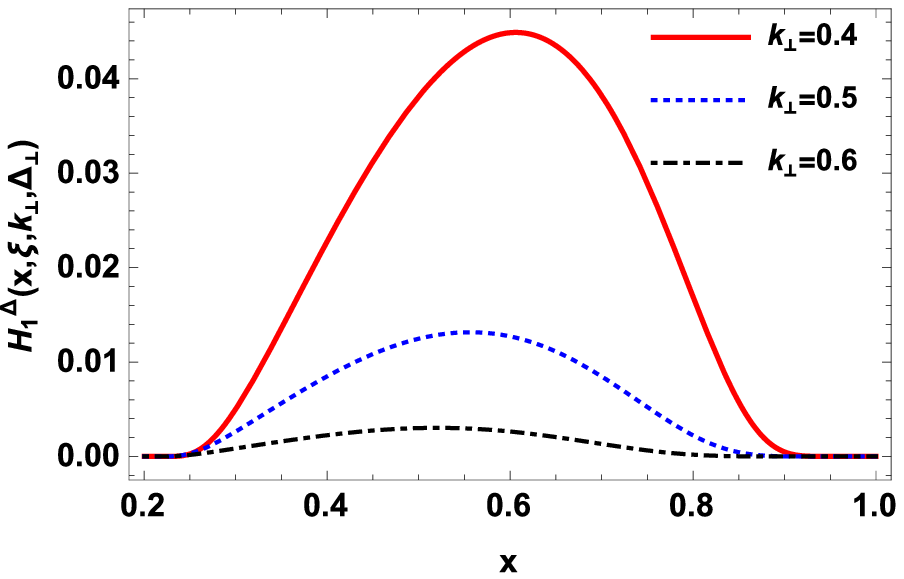}\hfill
(h)\includegraphics[width=.4\textwidth]{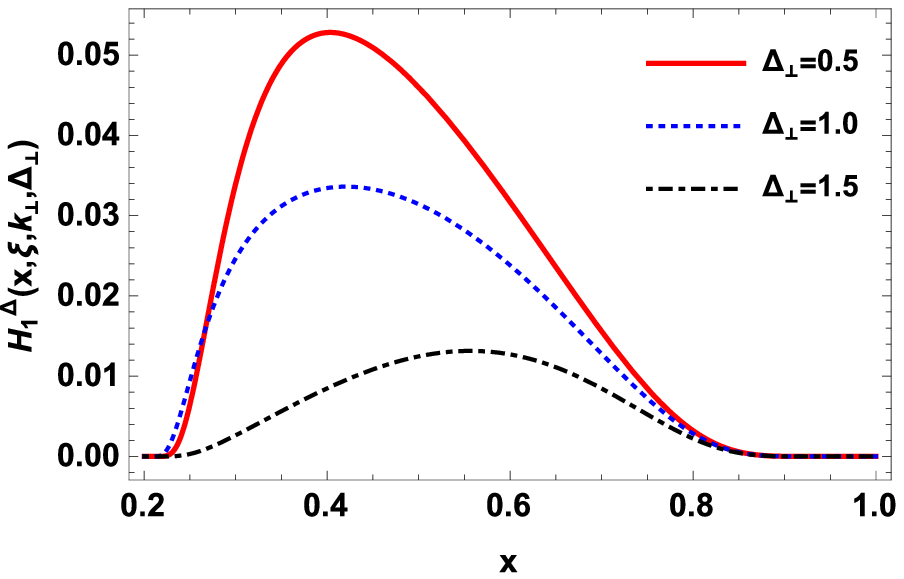}
\end{minipage}
\caption{The quark GTMDs $ F_{1} $, $ G_{1} $, $ H_{1}^{k} $ and $ H_{1}^{\Delta}$ for $ \xi = 0.1 $. The left panel shows the GTMDs  as a function of $ x $ with a fixed value of $ \bfd = 0.2 $ GeV and  different values of $ \bfk  (=0.4, 0.5, 0.6)$. The right panel shows the GTMDs as a function of $ x $  with a fixed value of $ \bfk = 0.5 $ GeV and for the different values of $ \bfd(=0.5, 1.0, 1.5)$.}
\label{plot_Forzeta_nonzero}
\end{figure*} 
\begin{figure*}
\centering
\small{(a)} \includegraphics[width=0.4 \columnwidth]{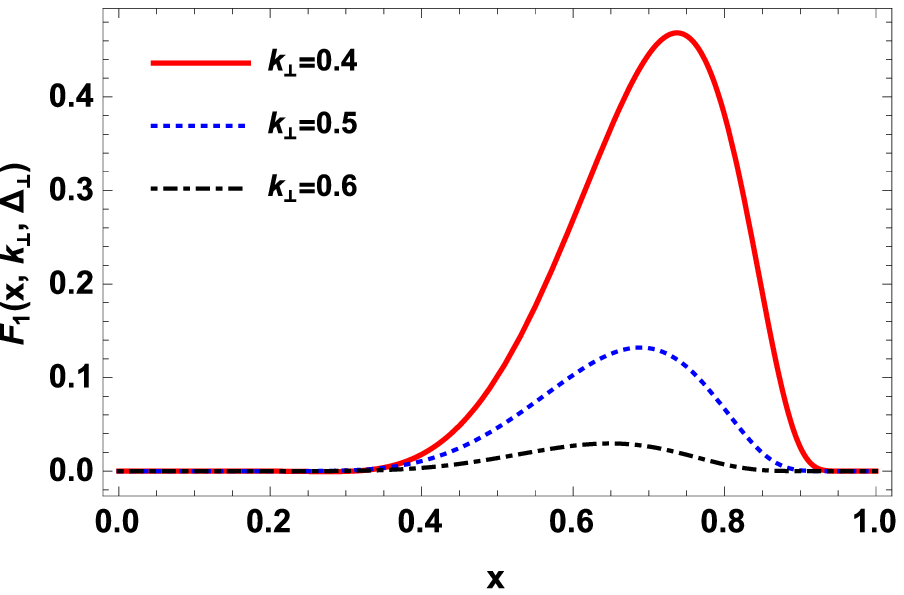}\hfill
\small{(b)} \includegraphics[width=0.4\columnwidth]{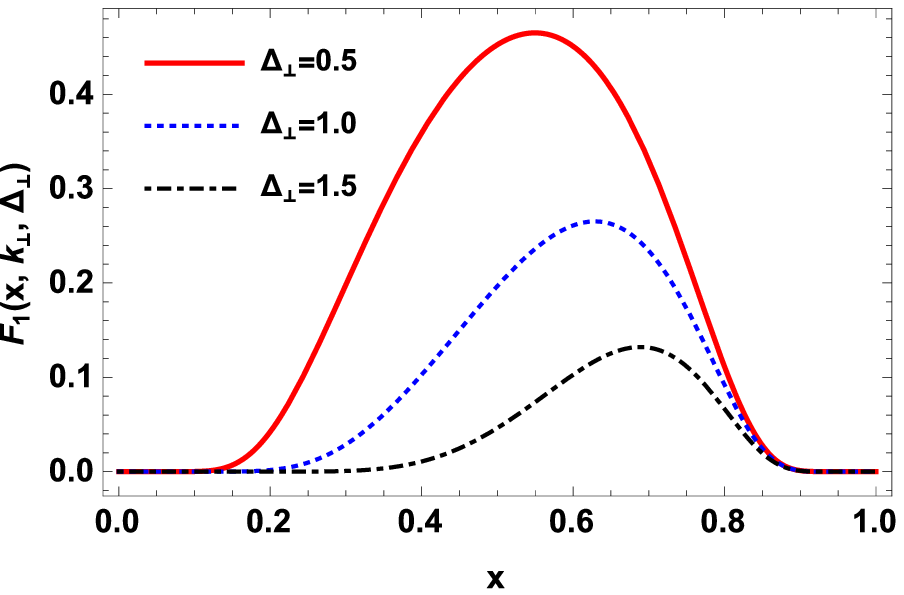}\\
\small{(c)} \includegraphics[width=0.4\columnwidth]{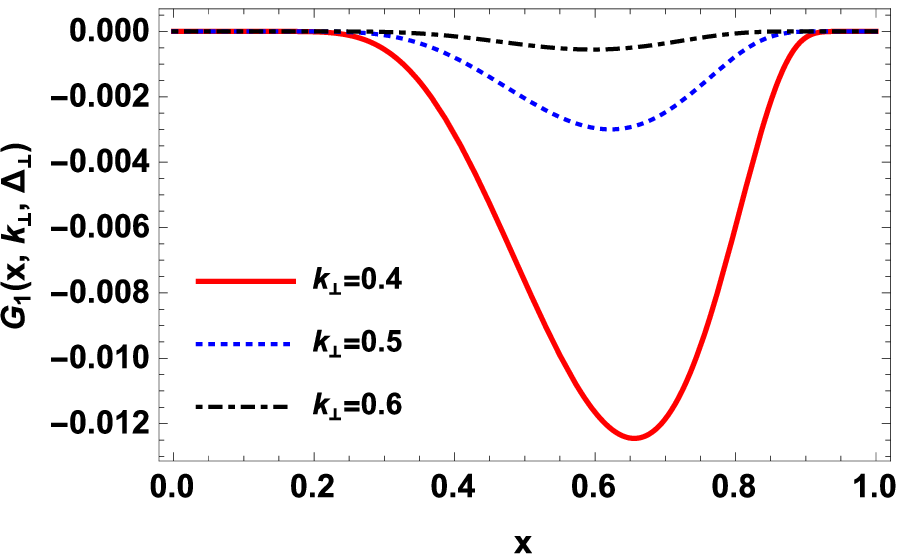}\hfill
\small{(d)} \includegraphics[width=0.4\columnwidth]{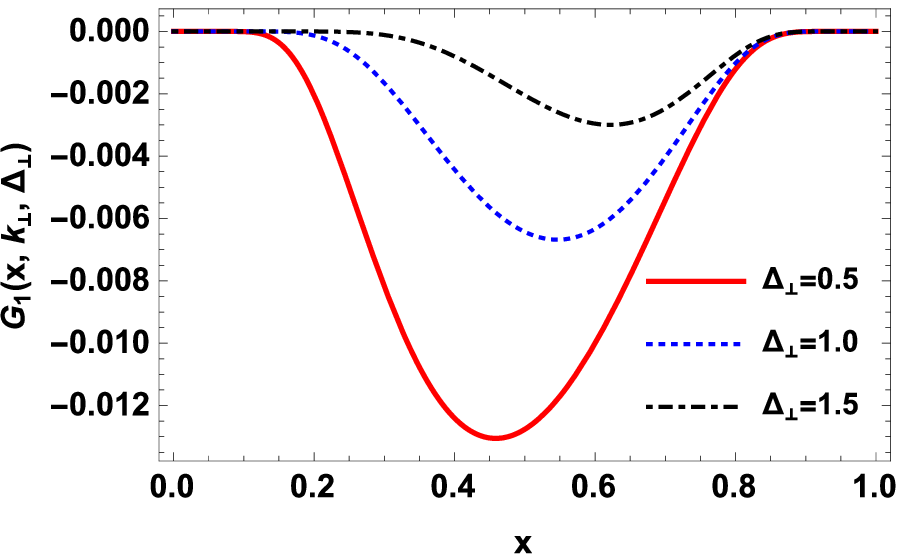}\\
\small{(e)} \includegraphics[width=0.4\columnwidth]{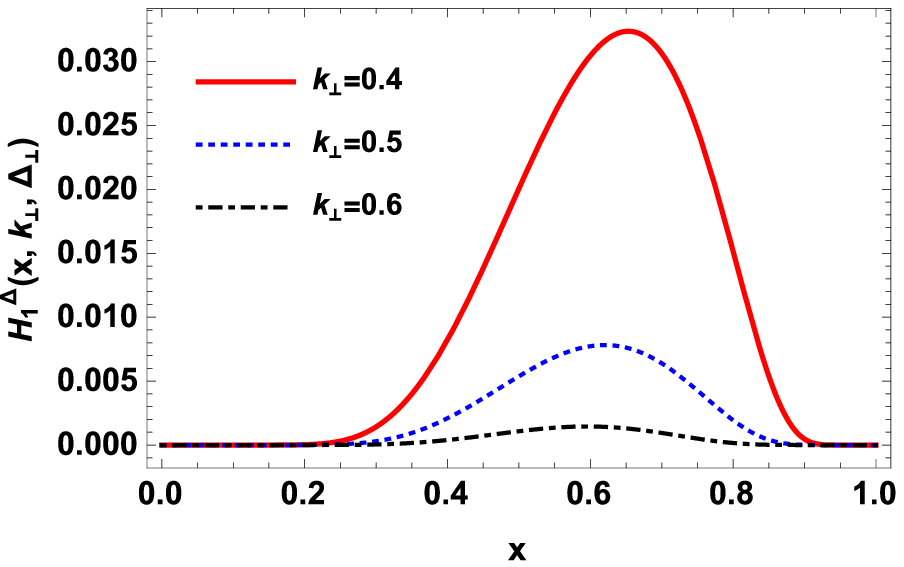}\hfill
\small{(f)} \includegraphics[width=0.4\columnwidth]{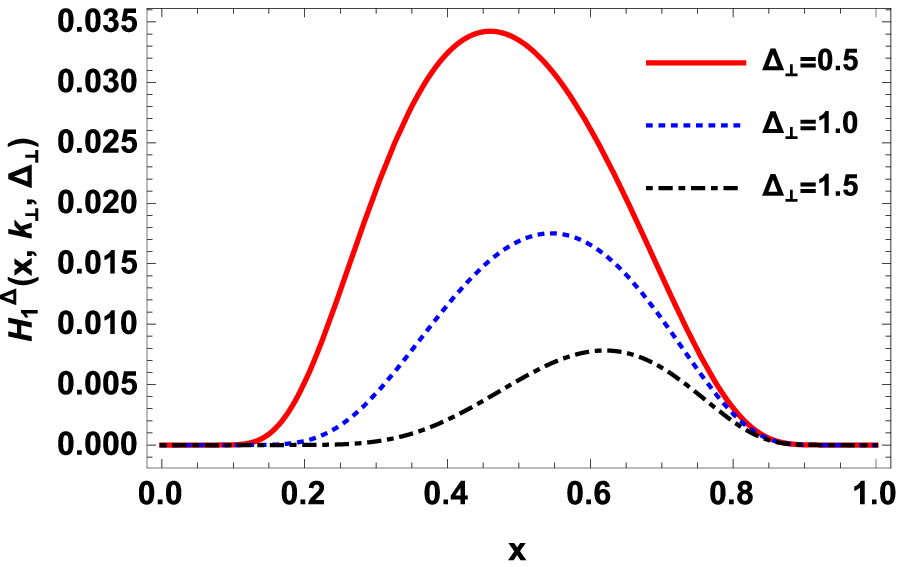}
\caption{The quark GTMDs $ F_{1} $, $ G_{1} $ and $ H_{1}^{\Delta}$ for $ \xi = 0 $. The left panel shows the GTMDs  as a function of $ x $ with a fixed value of $ \bfd = 0.2 $ GeV and  different values of $ \bfk  (=0.4, 0.5, 0.6)$. The right panel shows the GTMDs as a function of $ x $  with a fixed value of $ \bfk = 0.5 $ GeV for the different values of $ \bfd(=0.5, 1.0, 1.5)$.}
\label{plot_Forzetazero}
\end{figure*}

In Fig. \ref{plot_F1_G1}, we plot the quark GTMDs $ F_{1} (x,\,\xi ,\bfk, \bfd) $, $ G_{1}(x,\,\xi ,\bfk, \bfd) $, $ H_{1}^{k} (x,\,\xi ,\bfk, \bfd) $ and $ H_{1}^{\Delta} (x,\,\xi ,\bfk, \bfd) $ as a function of $ x $ by fixing the values of $ \bfk$ and $ \bfd$ as $ \bfk = 0.5 $ GeV and $ \bfd = 1.5 $ GeV. In all the plots we have presented the results for different values of $ \xi $ including  for $ \xi = 0 $.  We have observed that, in $ 0 < x < \xi $ region, all the distributions reduce to zero. Since the quark distributions  correspond to the DGLAP region $ \xi < x <1 $, all GTMDs vanish in non-DGLAP region ($ 0 < x < \xi $). 
It is clear from the plots that, with an increase in $ \xi $, the peaks of distributions $ F_{1} $, $ G_{1} $, $ H_{1}^{k} $ and $ H_{1}^{\Delta} $ shift towards higher values of $ x $. This clearly indicates that with increase in the longitudinal momentum transferred to pion, the longitudinal momentum carried by quark increases for fixed transverse momentum. For the case of GTMD $ F_{1} $ in Fig. \ref{plot_F1_G1} (a) it can be seen that the amplitude of $ F_{1} $ increases with the increase in $ \xi $ from $ 0 $ to $ 0.2 $ but with further increase in $ \xi $ ($ > 0.2 $), the amplitude decreases. This translates to the fact that for a fixed four momentum transfer, the distribution is maximum for smaller longitudinal momentum transfer $ \xi  < 0.2 $ and the distribution becomes smaller at larger longitudinal momentum transfer $ \xi  > 0.2 $. Integrating over $ \bfk $, the GTMD $ F_{1} $ can be reduced to the GPD $ H(x, \xi, \Delta^2) $ for nonzero skewness.
On the other hand, for the case of $ G_{1} $, $ H_{1}^{k} $ and $ H_{1}^{\Delta} $ in Fig. \ref{plot_F1_G1} (b), (c) and (d) respectively, the amplitude increases as $ \xi $ increases from value $ 0 $ to $ 0.5 $ but beyond $ 0.5 $, the amplitude decreases with increase in $ \xi $. The distribution amplitudes are maximum when longitudinal momentum transfer to pion is equally distributed between its constituent i.e quark and antiquark. For the case of $ H_{1}^{k} $ in Fig. \ref{plot_F1_G1} (c), the distribution vanishes for the case when $ \xi = 0  $. This is due to the term $ {\cal X} $ (Eq. \ref{XY}) which becomes zero leading to $ H_{1}^{k} $ GTMD being zero in this model for $ \xi = 0  $.  When we consider non-zero $ \xi $, $ H_{1}^{k} $ does not vanish and changes polarity from negative ($ - $) to positive ($ + $) along $ x $. The presence of $ {\cal X} $  in the expression of $ H_{1}^{k} $ brings the polarity change as $ {\cal X} $  changes sign for  $ x $ increasing for a fixed value of $ \xi $.   The behavior of $  G_{1} $ and $ H_{1}^{\Delta} $  (Fig. \ref{plot_F1_G1} (b) and \ref{plot_F1_G1} (d) respectively) is similar but the  polarities are opposite in both the cases and also they differ in terms of the magnitudes.

In Fig. \ref{plot_Forzeta_nonzero}, we have presented the results of quark GTMDs $ F_{1} $, $ G_{1} $, $ H_{1}^{k} $ and $ H_{1}^{\Delta}$ for $ \xi = 0.1 $. We have shown the plot for $ F_{1} $, $ G_{1} $, $ H_{1}^{k} $ and $ H_{1}^{\Delta}$ with a fixed value of $ \bfd = 0.2 $ GeV  and  different values of $ \bfk  (=0.4, 0.5, 0.6)$ in  Figs. \ref{plot_Forzeta_nonzero} (a), (c), (e) and (g) respectively. We observe that the amplitude of distributions decreases with increase in quark transverse momentum $ \bfk $ and peaks shift towards the lower value of $ x $. The implies that inside the pion, the probability of finding a quark with lower longitudinal momentum fraction and higher transverse momentum is less. In Figs. \ref{plot_Forzeta_nonzero} (b), (d), (f) and (h) respectively, we display the plots for $ F_{1} $, $ G_{1} $, $ H_{1}^{k} $ and $ H_{1}^{\Delta}$ with a fixed value of $ \bfk = 0.5 $ GeV and for the different values of $ \bfd(=0.5, 1.0, 1.5)$. We can see that with increase in $ \bfd $, the distribution peaks shift lower in amplitude and move towards the higher value of $ x $. As more transverse momentum transferred to the pion, the probability of finding a quark with higher longitudinal momentum decreases.
In Fig. \ref{plot_Forzeta_nonzero} (e) and \ref{plot_Forzeta_nonzero} (f), we can see that the $ H_{1}^{k} $ has a peak in negative direction for smaller values of $ x $ however,  in the higher $ x $ value region, the peak shifts in positive direction. This occurs due to the change in sign of $ {\cal X} $ factor which depends only on $ x $ and $ \xi $. Therefore, the polarity change of $ H_{1}^{k} $ occurs at $ x \approx 0.5 $ for different values of $ \bfk $  and $ \bfd $.

For $ \xi = 0 $ case, we have plotted the results of pion GTMDs $ F_{1} $, $ G_{1} $ and $ H_{1}^{\Delta}$  in Fig. \ref{plot_Forzetazero}. By comparison of Fig. \ref{plot_Forzeta_nonzero} and Fig. \ref{plot_Forzetazero}, we observe that the GTMDs $ F_{1} $, $ G_{1} $ and $ H_{1}^{\Delta}$ have similar variation along $ x $ for both $ \xi = 0 $ and  for $ \xi \neq 0 $ case. For $ \xi = 0 $ case, the distribution curves become narrower and amplitudes decrease as compared to $ \xi \neq 0 $ case. This implies that the quark distributions with no momentum transfer along longitudinal direction are smaller than the quark distributions with longitudinal momentum transfer. The GTMD $ H_{1}^{\Delta} $ vanishes in this model at $ \xi = 0 $ as $ {\cal{X}} $ goes zero.

\section{Summary}
In this paper, we have discussed the pion Wigner distributions which provide a multi-dimensional picture of pion. We have evaluated the Wigner distributions using the overlap representation of LFWFs obtained from light-front holographic model for pion.
We have studied the corresponding Wigner distribution of pion in transverse momentum space as well as in impact-parameter space for different quark polarizations i.e unpolarized, longitudinally polarized and transversely polarized inside an unpolarized pion. The distribution of unpolarized quark inside an unpolarized pion $ \rho_{UU} $ is spherically symmetric in impact-parameter space as well as in momentum space which implies that the probability of a quark to spin up is equivalent to the probability of quark to spin down. Due to advantaged direction of quark polarization, the distribution of longitudinally polarized quark inside an unpolarized pion $ \rho_{UL} $ has a dipolar distribution in both impact parameter space and in momentum space. To evaluate the distribution of transversely polarized quark inside an unpolarized pion $ \rho_{UT} $ we have considered the transverse polarization along x-axis. The direction of polarization must be perpendicular to the transverse co-ordinate in order to have a non-zero $ \rho_{UT} $. In impact-parameter space, $ \rho_{UT} $ has a dipolar structure but exhibits spherically symmetric behavior in transverse space which implies non-dependence of $ \rho_{UT} $ on transverse momentum direction. 

We have also analyzed the GTMDs $ F_{1} $, $ G_{1} $, $ H_{1}^{k} $ and $ H_{1}^{\Delta}$ of pion in this model for different values of skewness $ \xi $ as well as for $ \xi = 0 $ case. With skewness variation, we have observed the amplitude of all GTMDs going higher upto certain value of $ \xi $ after which they start decreasing. We have found a similar behavior for $ G_{1} $ and $ H_{1}^{\Delta}$ GTMDs but with opposite polarity. The $ H_{1}^{k} $ GTMD changes polarity with increase in longitudinal momentum for nonzero zeta in this model.   
For all distributions, the longitudinal momentum carried by quark increases with increase in longitudinal momentum transfer for fixed transverse momentum. For the transverse momentum variation, we have found the amplitude of distributions decreasing with the increase in quark transverse momentum for fixed longitudinal momentum and transverse momentum transfer. With the variation of transverse momentum transfer, we get information on the probability of finding a quark with higher longitudinal momentum and fixed transverse momentum decreases with increase in transverse momentum transfer.  
For $ \xi = 0 $ case, three twist-2 GTMDs $F_{1}$, $G_{1}^{k}$ and $H_{1}^{\Delta}$ remain for pion as $H_{1}^{\Delta}$ become zero in this model. This can perhaps be substantiated
further by future Drell-Yan measurements for the case of pion which would have important implications for the subtle features of the light-front holographic model.

\acknowledgements
H.D. would like to thank the Department of Science and Technology (Ref No. EMR/2017/001549) Government of India for financial support.

\end{document}